# A Constructive Algorithm to Prove P=NP

Duan Wen-Qi (wenqiduan@vip.126.com)

*College of Economics and Management, Zhejiang Normal University, Jinhua, 321004, China*

**Abstract:** After reducing the undirected Hamiltonian cycle problem into the TSP problem with cost 0 or 1, we developed an effective algorithm to compute the optimal tour of the transformed TSP. Our algorithm is described as a growth process: initially, constructing 4-vertexes optimal tour; next, one new vertex being added into the optimal tour in such a way to obtain the new optimal tour; then, repeating the previous step until all vertexes are included into the optimal tour. This paper has shown that our constructive algorithm can solve the undirected Hamiltonian cycle problem in polynomial time. According to Cook-Levin theorem, we argue that we have provided a constructive proof of **P=NP**.

This paper, taking Hamiltonian cycle as our object, wishes to develop a constructive algorithm to prove **P=NP**, which is one of the seven Millennium Prize Problems selected by the Clay Mathematics Institute, and is also a major unsolved problem in computer science. NP represents the class of questions for which there is no known way to find an answer quickly, but an answer can be verified in polynomial time. For the hardest NP problems (i.e., NP-complete problems), given an efficient algorithm for any one of them, we can find an efficient algorithm for all of them [1-4].

In terms of graph theory, a Hamiltonian cycle is a cycle in a graph that visits each vertex exactly once. For a given graph (whether directed or undirected), determining the existence of such cycles is the Hamiltonian cycle problem, which is NP-complete problem [3]. According to the Cook-Levin theorem, if we solve the undirected Hamiltonian cycle problem in polynomial time, we also provides a constructive proof of **P=NP**. Following is our constructive algorithm. We first reduce the undirected Hamiltonian cycle problem into a special TSP problem. Let $G = (V, E)$ be an instance of the undirected Hamiltonian cycle problem. Consider a complete graph $\overline{G} = (V, \overline{E})$ where $\overline{E} = \{(u,v) \mid u,v \text{ in } V \text{ and } u \neq v\}$. Assign a cost to each edge in $\overline{E}$ as follows: (1) $c(u,v) = 0$ if $(u,v)$ in $E$; (2) $c(u,v) = 1$ if $(u,v)$ not in $E$. Evidently, if the total cost of the optimal tour of $\overline{G}$ is 0, then graph $G$ must have at least one Hamiltonian cycle. Otherwise, there is no Hamiltonian cycle in graph $G$. Therefore, we will prove **P=NP** provided that we can develop a deterministic algorithm to solve the transformed TSP problem $\overline{G}$ in polynomial time.

The number of vertexes in $\overline{G}$ is $N = |V|$, and $v_i (i = 1,2,\ldots,N)$ denotes the $i$th vertex. Let $C_m^*$ be the total cost of the optimal tour, which corresponds to the $m$-TSP composed of $v_1$, $v_2$, …, and $v_m$. When $C_m^* \geq 1$, if any $c(v_i, v_j) = 1$ and the edge $(v_i, v_j)$ can appear in any optimal tour (whatever one or multiples optimal tours exist) of the $m$-TSP, we call this kind of edges as optimizing edges (denoted as $(v_i^*, v_j^*)$) because of their key role playing in our algorithm. Let $H_m = \{(v_i^*, v_j^*) \mid i, j \in [1, m]\}$, which contains all optimizing edges. Evidently, whether an edge is optimizing edge depends on the specific $m$-TSP. $H_m$ will change if the value of $m$ is changed. For each optimizing edge, we construct an optimal tour which must contain that edge. Therefore, we will construct $K_m = |H_m|$ optimal tours for the $m$-TSP. Let $\omega_m^i$ ($i = 1,2,\ldots,K_m$) be the $i$th optimal tour which contains the $i$th optimizing edge in $H_m$, and $\Psi_m = \{\omega_m^i \mid i = 1,2,\ldots,K_m\}$ be the set of $K_m$ optimal tours. When $C_m^* = 0$, it will happen $H_m = \varnothing$, $K_m = 0$, and $\Psi_m = \varnothing$. In order to keep consistence, for those $m$-TSP problems with $C_m^* = 0$, we define $\overline{H}_m$ as the set of (1) edges must be of cost 0 and appear in any optimal tour whose total cost is 0, and (2) edges must be of cost 1 and appear in any tour whose total cost is 1. Considering that edges in $\overline{H}_m$ will play the same role as edges in $H_m$, we also call them as optimizing edges, call tours containing optimizing edge with cost 0 as optimal tours and tours containing one optimizing edge with cost 1 as "optimal tours". We define $\overline{K}_m = |\overline{H}_m|$ as the number of edges in $\overline{H}_m$, $\overline{\omega}_m^i$ ($i = 1,2,\ldots,\overline{K}_m$) be the $i$th optimal tour or "optimal tour" which contains the $i$th optimizing edge, and $\overline{\Psi}_m = \{\overline{\omega}_m^i \mid i = 1,2,\ldots,\overline{K}_m\}$ be the set of tours which will be used in vertex growth.

When the new vertex $v_{m+1}$ is added into the $m$-TSP, the total cost of the optimal tour of $m+1$-TSP depends on $C_m^*$, $H_m$ ($\overline{H}_m$), $D_{m+1}^*$ = Min $\{d_{i,j}^{m+1}=c(v_{m+1},v_i)+c(v_{m+1},v_j)\,|\,i\neq j\in[1,m]\}$, $\Omega_{m+1}=\{(v_i,v_j)\,|\,d_{i,j}^{m+1}=D_{m+1}^*,i\neq j\in[1,m]\}$, and $\Omega_{m+1}^s=\{(v_i,v_j)\,|\,d_{i,j}^{m+1}=D_{m+1}^*+1,i\neq j\in[1,m]\}$. Table 1 shows in detail how $C_{m+1}^*$ depends on the specific value of $C_m^*$, $D_{m+1}^*$, $H_m$ ($\overline{H}_m$), $\Omega_{m+1}$, and $\Omega_{m+1}^s$. It is worth noting that $C_N^*$ is the total cost of the optimal tour of $\overline{G}$. If $C_N^*=0$, it implies that there is at least one Hamiltonian cycle in graph $G$; otherwise there is no Hamiltonian cycle in graph $G$. Evidently, $C_{m+1}^*$, $D_{m+1}^*$, $\Omega_{m+1}$, and $\Omega_{m+1}^s$ are very easy to compute if we know $C_m^*$ and $H_m$ ($\overline{H}_m$). Therefore, we will solve the transformed TSP problem $\overline{G}$ in polynomial time provided that our algorithm can construct $H_m$ ($\overline{H}_m$) and $\Psi_m$ ($\overline{\Psi}_m$) correctly from $m=4$ to $m=N$ in polynomial time, which will be proved in the latter.

Table 1 The total cost of the optimal tour with $m+1$ vertexes

| $D_{m+1}^*$ | $C_m^*=0$ | $C_m^*\geq 1$ |
|---|---|---|
| 0 | if $(\overline{H}_m\cap\Omega_{m+1})\neq\varnothing$ then $C_{m+1}^*=0$ | if $(H_m\cap\Omega_{m+1})\neq\varnothing$ then $C_{m+1}^*=C_m^*-1$ |
| 0 | if $(\overline{H}_m\cap\Omega_{m+1})=\varnothing$ then $C_{m+1}^*=1$ | if $(H_m\cap\Omega_{m+1})=\varnothing$ and $(H_m\cap\Omega_{m+1}^s)\neq\varnothing$ then $C_{m+1}^*=C_m^*$ |
| 0 | | if $(H_m\cap\Omega_{m+1})=\varnothing$ and $(H_m\cap\Omega_{m+1}^s)=\varnothing$ then $C_{m+1}^*=C_m^*+1$ |
| 1 | $C_{m+1}^*=1$ | if $(H_m\cap\Omega_{m+1})\neq\varnothing$ then $C_{m+1}^*=C_m^*$ |
| 1 | | if $(H_m\cap\Omega_{m+1})=\varnothing$ then $C_{m+1}^*=C_m^*+1$ |
| 2 | $C_{m+1}^*=2$ | $C_{m+1}^*=C_m^*+1$ |

After introducing our general idea, we now present the detailed process to solve the TSP problem.

**Step one**: starting from $m=4$, picking out four vertices and making sure that the total cost of the optimal tour of the 4-TSP is larger than 0, i.e. $1\leq C_4^*<4$; computing $C_4^*$, $H_4$, and $\Psi_4$. If the total cost of the optimal tours of all 4-TSP in $\overline{G}$ is 0, then there must exist multiple Hamiltonian cycles in $G$.

**Step two**: adding the new vertex $v_{m+1}$: as $C_m^*$ and $H_m$ ($\overline{H}_m$) already known, computing $D_{m+1}^*$, $\Omega_{m+1}$, $\Omega_{m+1}^s$ (if necessary), and $C_{m+1}^*$ based on Table 1, and obtaining one optimal tour $\omega_{m+1}^*$ with $\Psi_m$ ($\overline{\Psi}_m$). The optimal tour $\omega_{m+1}^*$ is constructed in such a way that the lowest total cost $C_{m+1}^*$ in table 1 will be realized. Following we describe in detail how to construct $\omega_{m+1}^*$ under different situations. **Case one**: $D_{m+1}^*=2$. In this case, we first (1) randomly choose one edge (if $C_m^*=0$), or (2) choose one optimizing edge (if $C_m^*\geq 1$) from one of the optimal tour(s) in $\overline{\Psi}_m$ (or $\Psi_m$), then connect $v_{m+1}$ to the two vertices of the previous chosen edge and delete that edge. **Case two**: $D_{m+1}^*=1$. In this case, there must exist one, but only one vertex $v_l$ which makes $(v_{m+1},v_l)=0$. We pick out $v_l$ and (1) find out the optimal tour which contains an edge $(v_l,v_i)=1$ from $\Psi_m$ if $(H_m\cap\Omega_{m+1})\neq\varnothing$ or (2) pick out another vertex $v_j$ provided that $(v_l,v_j)$ appears in one optimal tour if $(H_m\cap\Omega_{m+1})=\varnothing$ or $C_m^*=0$, then connect two edges $(v_{m+1},v_l)$ and $(v_{m+1},v_i)$ (or $(v_{m+1},v_j)$), and delete the edge $(v_l,v_i$ (or $v_j))$ from the corresponding optimal tour. **Case three**: $D_{m+1}^*=0$. In this case, if $C_m^*=0$ and $(\overline{H}_m\cap\Omega_{m+1})\neq\varnothing$, we first find out an edge $(v_i,v_j)$ so as to $(v_{m+1},v_i)=0$ and $(v_{m+1},v_j)=0$ from $\overline{\Psi}_m$, then pick out that tour which contains $(v_i,v_j)$, add two edges $(v_{m+1},v_i)$ and $(v_{m+1},v_j)$ into, and delete the edge $(v_i,v_j)$ from it. If $C_m^*=0$ and $(\overline{H}_m\cap\Omega_{m+1})=\varnothing$, we randomly pick out one optimal tour from $\overline{\Psi}_m$, and connect the new vertex $v_{m+1}$ to the vertex $v_i$ which satisfies with $(v_{m+1},v_i)=0$ and $v_i$'s adjacent vertex $v_j$, then delete the edge $(v_i,v_j)$ from that optimal tour. If $C_m^*\geq 1$ and $(H_m\cap\Omega_{m+1})\neq\varnothing$, we first find out one edge $(v_i,v_j)$ that satisfies with $(v_{m+1},v_i)=0$ and $(v_{m+1},v_j)=0$ from $\Psi_m$, then pick out the optimal tour which contains the edge $(v_i,v_j)$, add the edges $(v_{m+1},v_i)$ and $(v_{m+1},v_j)$ into, and delete $(v_i,v_j)$ from that optimal tour. Lastly, considering situation $(H_m\cap\Omega_{m+1})=\varnothing$ and $C_m^*\geq 1$ ($(H_m\cap\Omega_{m+1}^s)\neq\varnothing$ or $(H_m\cap\Omega_{m+1}^s)=\varnothing$), it is very similar to the **Case two** with $C_m^*\geq 1$ if we replace the symbol $\Omega_{m+1}^s$ with $\Omega_{m+1}$. Hence, we can construct the optimal tour with the procedure used in **Case two**.

**Step three**: constructing $H_{m+1}$ ($\overline{H}_{m+1}$) and $\Psi_{m+1}$ ($\overline{\Psi}_{m+1}$) based on the optimal tour $\omega_{m+1}^*$. If $C_{m+1}^*\geq 1$, we use **Optimizing Edge Replacing** (OER) moves to construct $H_{m+1}$ and $\Psi_{m+1}$. If $C_{m+1}^*=0$, we use **Modified Optimizing Edge Replacing** (MOER) moves to construct $\overline{H}_{m+1}$ and $\overline{\Psi}_{m+1}$. The implementation of OER and MOER will be illustrated in the latter.

**Step four**: repeating Step two and Step three until $m=N$. After obtained $C_N^*$, $H_N$ ($\overline{H}_N$), and $\Psi_N$ ($\overline{\Psi}_N$), we complete our computing procedure for solving the undirected Hamiltonian cycle problem.

The key idea of **OER** is to find out all new optimizing edges which can replace those optimizing edge(s) already existed in $\omega_{m+1}^*$. To implement OER move, we first start from any out-optimizing edge $(v_i, v_j) = 1$ in $\omega_{m+1}^*$, then exhaustive search and perform sequential 2-opt and 3-opt moves like that in Lin–Kernighan heuristic [5-6], and perform double-bridge non-sequential move which is shown in Fig. 4 of Ref. [6], so as to find out at least one new optimizing edge. It is worth noting that all sequential and non-sequential moves in OER must not change the total cost of optimal tour(s). After each move, at least one new optimizing edge is introduced into the optimal tour, which will be used as out-optimizing edge in latter OER moves. We iteratively start from one optimizing edge to perform succession OER moves until we cannot find out any new optimizing edge. It is easy to see that the complexity of performing OER move is $O(5)$ in worst case, which implies our computation can be finished in polynomial time.

Different from $H_{m+1}$ only containing edges with cost 1, $\overline{H}_{m+1}$ contains not only edges with cost 0 but also edges with cost 1. Correspondingly, $\overline{\Psi}_{m+1}$ contains two kinds of tours. One is the optimal tours whose total cost must be 0, and the other is the "optimal tours" whose total cost must be 1. MOER move is much like OER move, i.e., replacing out-edges with other in-edges by performing sequential 2-opt, 3-opt moves, and double-bridge non-sequential move. However, there are major differences between MOER and OER. There are four kinds of OER moves in MOER. The first one is zero-OER, which is used to find out optimizing edges with cost 0 in $\overline{H}_{m+1}$ and optimal tours in $\overline{\Psi}_{m+1}$. Evidently, both in-edges and out-edges all are optimizing edges. When we perform zero-OER moves, the total cost of the optimal tour must keep unchanged. The second one is add-one-OER, in which the total cost of the original tour increases to 1 after performing any move, whatever sequential 2-opt, 3-opt moves, or double-bridge non-sequential move. After an add-one-OER, only in-edges with cost 1 (being taken as optimizing edges) and an "optimal tour" contained that will be added into $\overline{H}_{m+1}$ and $\overline{\Psi}_{m+1}$, respectively. The third one is minus-one-OER, which is the inverse move of add-one-OER. The "optimal tour" will become an optimal tour and the total cost of the "optimal tour" will decrease to 0 after performing a move. Evidently, in minus-one-OER, all in-edges must be of cost 0, and all of them are taken as optimizing edges and added into $\overline{H}_{m+1}$. The fourth one is the same as OER move, which does not change the total cost of the original "optimal tour", and only the in-edges with cost 1 will be taken as new optimizing edges. As described in the above, based on an optimal tour, we can perform zero-OER or add-one-OER; we can implement minus-one-OER and OER move based on an "optimal tour". For the given optimal tour $\omega_{m+1}^*$, if $C_{m+1}^* = 0$, all edges in it are optimizing edges. We start from these initially optimizing edges to exhaustively perform the above four moves so as to find out new optimizing edges. By iteration, we will find out all optimizing edges, and construct all "optimal tours" and optimal tours. Similar to OER, the complexity of MOER is $O(5)$ in worst case, which means that the whole computation to construct $\overline{H}_m$ and $\overline{\Psi}_m$ can be finished in polynomial time.

As described in the above, our algorithm can solve the undirected Hamiltonian cycle problem in polynomial time completely provided that we can find out all optimizing edges in $H_{m+1}$ ($\overline{H}_{m+1}$) based on a given optimal tour $\omega_{m+1}^*$. Following we illustrate why we can find out all optimizing edges in $H_{m+1}$ ($\overline{H}_{m+1}$) by applying OER (MOER) moves based on a give optimal tour. For any vertex $v_i$, if there exists at least one edge $(v_i, v_j) \in H_{m+1}(\overline{H}_{m+1})$, we call it as optimizing vertex, otherwise as non-optimizing vertex. For any optimizing edge $(v_i, v_j)$ with the vertex $v_i$, its each connected edge $(v_i, v_j')$ with cost 0 must correspond to one optimizing edge $(v_j, v_i')$, which make us obtain a new optimal tour (or "optimal tour") with new optimizing edge only by a simple OER (MOER) move. Conversely, if $C^*_{m+1} = 0$, each optimizing edge in $\overline{H}_{m+1}$ must have at least one edge with zero cost corresponding to it. Evidently, for $C^*_{m+1} = 0$ and $\overline{H}_{m+1}$, all vertices are optimizing vertices and each vertex connects to at least two optimizing edges. Therefore, MOER move can exploit all zero cost edges of each vertex and find out all optimizing edges.

For $C^*_{m+1} \geq 1$ and $H_{m+1}$, we define $V^{opt}_{m+1} = \{v_i | \exists (v_i, v_j) \in H_{m+1}\}$ as the set of optimizing vertices and $\overline{V}^{opt}_{m+1} = \{v_i | \forall (v_i, v_j) \notin H_{m+1}\}$ as the set of non-optimizing vertices. For any two vertices $v_i \in V^{opt}_{m+1}$ and $v_j \in V^{opt}_{m+1}$, if $(v_i, v_j) \in H_{m+1}$, then there is an edge between $v_i$ and $v_j$ which corresponds to the optimizing edge $(v_i, v_j)$. By this way, we can obtain a graph $G^{opt}_{m+1}$. For any two vertices $v_i$ and $v_j$ in $V^{opt}_{m+1}$ and $(v_i, v_j) \neq 1$, if there exists a path between them, we can apply succession OER moves to obtain another optimal tour containing $(v_j, v_l) \in H_{m+1}$ from one optimal tour containing $(v_i, v_k) \in H_{m+1}$. Therefore, provided that $G^{opt}_{m+1}$ is a connected graph, we can find out all optimizing edges from the seed optimizing edge(s) in $\omega^*_{m+1}$ by applying succession OER moves. Following we apply inductive method to prove $G^{opt}_{m+1}$ is a connected graph.

For $m+1 = 4$, without doubt, we can obtain $H_4$ and $\Psi_4$ correctly by applying OER moves based on one optimal tour $\omega^*_4$. Now, we assume that we can obtain $H_m$ and $\Psi_m$ by applying OER moves based on a optimal tour $\omega^*_m$. Evidently, $G^{opt}_m$ should be a connected graph under the previous assumption. When the new edges $(v_{m+1}, v_i)$ and $(v_{m+1}, v_j)$ are added into and the optimal tour $\omega^*_{m+1}$ is given, we examine whether $G^{opt}_{m+1}$ is a connected graph. If $C^*_m \geq 1$ and $c(v_{m+1}, v_i) + c(v_{m+1}, v_j) = 0$, then $H_{m+1} \subseteq H_m$ and $V^{opt}_{m+1} \subseteq V^{opt}_m$. Being a subgraph of $G^{opt}_m$, $G^{opt}_{m+1}$ must be a connected graph. If $c(v_{m+1}, v_i) + c(v_{m+1}, v_j) = 2$, then $C^*_{m+1} \geq 2$ and $v_{m+1}$ can connect to any vertex in $\omega^*_m$ if $C^*_m = 0$ (or connect to two optimizing vertices $v_i$ and $v_j$ satisfied with $(v_i, v_j) = 1$). It is easy to see that $G^{opt}_{m+1}$ is a connected graph. If $c(v_{m+1}, v_i) + c(v_{m+1}, v_j) = 1$, without lost generality, let $c(v_{m+1}, v_i) = 0$ and $c(v_{m+1}, v_j) = 1$. In this case, under the situation of $C^*_m \geq 1$ and $C^*_{m+1} \geq 1$, as we can see from Table 1 and Step two, $v_{m+1}$ as an optimizing vertex, has connected to $G^{opt}_m$ through the edge $(v_{m+1}, v_j)$ because of $(v_i, v_j) = 1$ or we can builds the connection by implementing only one OER move between $(v_{m+1}, v_j)$ and another optimizing edge in $\omega^*_{m+1}$ (if $(v_i, v_j) = 0$, then $C^*_{m+1} \geq 2$). Under the situation of $C^*_m = 0$ and $C^*_{m+1} = 1$, all optimizing vertices can be connected to $v_{m+1}$, at most applying 2 OER moves. Hence, all optimizing vertices are composed of a connected graph. As analyzed in the above, in whatever situation, $G^{opt}_{m+1}$ must be a connected graph if $G^{opt}_m$ is already a connected graph. Therefore, we have proved that applying OER move can find out all optimizing edges in $H_N$ for any $N$.

As shown in the previous paragraph, our constructive algorithm can solve one of the most famous NP-complete problems — the undirected Hamiltonian cycle problem in polynomial time. According to the Cook–Levin theorem, we conclude that we have provided a constructive proof of **P=NP**.